%% file: optical_bind.tex
\documentstyle[aps,prb,graphics,multicol,epsfig]{revtex}

\def\bs#1{\bbox{#1}}
\def\ve#1{\bs{\mathbf{#1}}}
\newcommand{\be}{\begin{eqnarray}}
\newcommand{\ee}{\end{eqnarray}}

\begin{document}

\draft \title{Optical binding of particles with or without the presence of a
flat dielectric surface} \author{P. C. Chaumet and
M. Nieto-Vesperinas}

\address{Instituto de Ciencia de Materiales de Madrid, Consejo
Superior de Investigaciones Cientificas, Campus de Cantoblanco Madrid
28049, Spain}

\maketitle

\begin{abstract}

Optical fields can induce forces between microscopic objects, thus
giving rise to new structures of matter.  We study theoretically these
optical forces between two spheres, either isolated in water, or in
presence of a flat dielectric surface. We observe different behavior
in the binding force between particles at large and at small distances
(in comparison with the wavelength) from each other.  This is due to
the great contribution of evanescent waves at short distances. We
analyze how the optical binding depends of the size of the particles,
the material composing them, the wavelength and, above all, on the
polarization of the incident beam.  We also show that depending on the
polarization, the force between small particles at small distances
changes its sign.  Finally, the presence of a substrate surface is
analyzed showing that it only slightly changes the magnitudes of the
forces, but not their qualitative nature, except when one employs
total internal reflection, case in which the particles are induced to
move together along the surface.

\end{abstract}



\begin{multicols}{2}

\section{Introduction}

Some time ago, it was demonstrated that optical fields can produce
forces on neutral particles,~\cite{ashkin69,ashkin70} since then this
mechanical action has been used in optical tweezers~\cite{ashkin86}
and more recently in optical force microscopy,~\cite{clapp,dogarin} as
well as in manipulating molecules~\cite{ashkin87} and dielectric
spheres.~\cite{collins,gauthier,taguchi} In addition, the possibility
of binding objects~\cite{burns} through optical forces and thus create
microstructures, either in or off resonant
conditions,~\cite{anto1,anto2,bayer} were pointed out.

In this paper we wish to undertake a detailed study on optical forces
on neutral particles, based on a rigorous analysis that we have
carried out~\cite{prbchaumetnieto,oplchaumetnieto,prbchaumetnieto2} in
a full three-dimensional configuration by using the coupled dipole
method (CDM) of Purcell and Pennypacker~\cite{purcell} via the
Maxwell's stress tensor.~\cite{stratton} Specifically, we study the
forces induced by light between two spheres, either isolated in
solution, or in the presence of a flat dielectric surface. We shall
monitor the nature, either attractive or repulsive, of the light
induced force between the spheres, according to the wavelength,
polarization of the incident wave, and the size and composition of the
spheres.

In Section~\ref{method} we outline the calculation method employed to
determine the optical binding forces, then, in Section~\ref{results}
we present results for spheres either isolated in water
(\ref{resultsA}) or suspended in this liquid in the presence of a flat
dielectric interface (\ref{resultsB}).

\section{Method used for computing the optical binding}\label{method}

In a previous article~\cite{prbchaumetnieto} we showed the possibility
to compute the optical forces on a sphere with the coupled dipole
method (CDM).~\cite{purcell} For the computation of the optical
binding between particles, we now use the same procedure, thus we
shall next only outline the main equations and the changes introduced
in them to address the presence of multiple objects.

Let $K$ objects be above a flat dielectric surface.  Each object is
discretized into $N_k$ subunits, with $k=1,\ldots,K$. Following the
procedure of Ref.~[\ref{purcell}], the field at the $(i,k)th$ subunit,
namely, the $ith$ subunit of the $kth$ object, can be written
as: \be
\label{dipi} \ve{E}(\ve{r}^k_i,\omega) & = &
\ve{E}_0(\ve{r}^k_i,\omega) + \sum_{l=1}^{K} \sum_{j=1}^{N_l} [
\ve{S}(\ve{r}^k_i,\ve{r}^l_j,\omega)\nonumber\\ & + &
\ve{T}(\ve{r}^k_i,\ve{r}^l_j,\omega)] \alpha^l_j
\ve{E}(\ve{r}^l_j,\omega). \ee where $\alpha^l_j$ is the
polarizability of the $(j,l)$ subunit, $\ve{T}$ is the linear response
to a dipole in free space,~\cite{jackson} and $\ve{S}$ represents the
linear response of a dipole in the presence of a
surface.~\cite{agarwal,rahmani} The value of the electric field at
each subunit position is obtained by solving the linear system
Eq.~(\ref{dipi}) written for all subunits, so that the size of the
system to solve is $\prod_{k=1}^K N_k$. Once the electric field is
obtained, the component of the total averaged force on the $(i,k)th$
subunit can be deduced from both the field and its derivative at its
position $\ve{r}^k_i$:~\cite{oplchaumetnieto} \be\label{forcec}
F_u(\ve{r}^k_i)=(1/2)\Re e\sum_{v=1}^{3}\nonumber\\
\left(p_v(\ve{r}^k_i,\omega) \frac{\partial
E_v^*(\ve{r}^k_i,\omega)}{\partial u}\right), \text{ ($u$=1, 2, 3).}
\ee where $u$, $v$, stand for either $x,y,z$, and
$\ve{p}(\ve{r}^k_i,\omega)$ is the electric dipole of the $(i,k)th$
subunit due to the incident field and all the other subunits. Notice,
that the derivative of the field can be obtained from the derivative
of Eq.~(\ref{dipi}).~\cite{prbchaumetnieto} Then the following
relation can be written: \be\ve{F}^k=\sum_{i=1}^{N_k}
\ve{F}(\ve{r}^k_i)\ee where $\ve{F}^k$ is the total force on the $kth$
object due to both the incident field and the multiple interaction
with the surface and the other $(K-1)th$ objects. If the object $kth$
is a sphere small compared to the wavelength, the dipole approximation
can be done, hence $N_k=1$. We also remark that, in what follows, when
we represent the normalized force, this means
$\ve{F}/(4\pi\varepsilon_0 |\ve{E}_i|^2)$ where $\varepsilon_0$ is the
permittivity of vacuum and $|\ve{E}_i|^2$ denotes the intensity of the
incident beam.

\section{Results and Discussion}\label{results}

In Fig.~1 we represent the more complex geometry that we shall
consider in this work. Two spheres (either dielectric or metallic) are
embedded in water ($\varepsilon_w=1.69$).  Illumination with an
incident plane wave takes place in the $XZ$ plane at an angle $\theta$
of incidence. When a dielectric flat surface at $z=0$ is used, we
consider it separating glass ($\varepsilon_s=2.25$) at $z<0$ from
water ($z>0$).

\begin{figure}[H]  
\begin{center}
\resizebox{80mm}{!}{\input{fig1.pstex_t}}
\end{center}
\caption{The most complex geometry considered in this paper: two
spheres of radius $r$ on a dielectric flat surface. The spheres are
embedded in water with $\varepsilon_w=1.69$, and the relative
permittivity of the surface is $\varepsilon_s=2.25$. The incident wave
vector $\ve{k}_0$ is in the $XZ$ plane, and $\theta$ is the angle of
incidence.}
\end{figure}
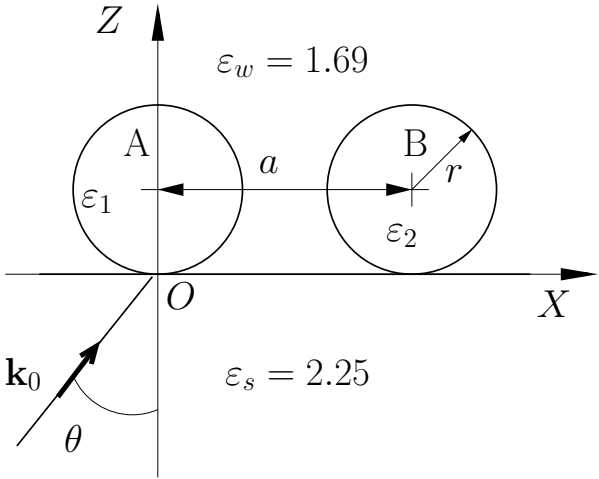

\subsection{Particles in water}\label{resultsA}

In this section we do not address yet the presence of the surface
($\varepsilon_s=\varepsilon_w=1.69$),
i.e. $\ve{S}(\ve{r}^k_i,\ve{r}^l_j,\omega)=0$ in Eq.~(\ref{dipi}), and
the angle of incidence is $\theta=0^{\circ}$. Even in the absence of
surface, we make reference to the polarization and thus we shall
always use the terms $p$-polarization and $s$-polarization when the
electric field vector is in the $XZ$ plane and along the $Y$-axis,
respectively.

\begin{figure}[H]  
\begin{center}
\includegraphics*[draft=false,width=80mm]{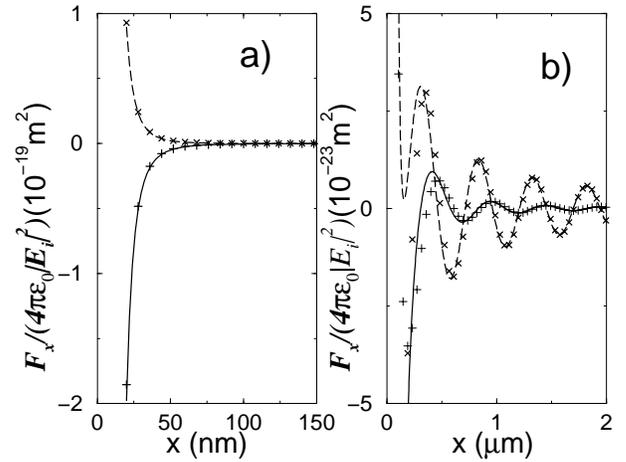}
\end{center}
\caption{Normalized force in the $X$-direction on sphere B versus
 distance $x$ between the centers of the spheres.  Both spheres are of
 glass ($\varepsilon_1=\varepsilon_2=2.25$), with $r=10$nm. The angle
 of incidence of the illuminating plane wave is $\theta=0^{\circ}$ and
 the wavelength $\lambda=632.8$nm in vacuum. The full line corresponds
 to $p$-polarization, and the dashed line represents
 $s$-polarization. a) Force for short distances between the spheres,
 the symbol + ($\times$) corresponds to the values from the
 non-retarded approximation for $p$-polarization
 ($s$-polarization). b) Force in far field, the symbol + ($\times$)
 represents the values from the non-retarded approximation in far
 field for $p$-polarization ($s$-polarization).}
\end{figure}

We begin with the most simple case, i.e., the radius $r$ of the two
particles is small compared to the wavelength employed. As previously
said, we then use the dipole approximation. We study, first, the case
of two identical spheres with $\varepsilon_1=\varepsilon_2=2.25$,
radius $r=$10nm at a wavelength $\lambda=$632.8nm in vacuum. Fig.~2
represents the force along the $X$-direction on the sphere B at
different positions of this sphere on the $X$-axis. The sphere A
remaining fixed. We have only plotted the force exerted on sphere B,
since by symmetry, the force along the $X$-axis on sphere A is the
opposite to that on B. We observe two facts: first, the oscillation of
the force when the spheres are far from each other, and, second, the
strong force, either attractive or repulsive, when the spheres are
very close from each other, depending on the polarization. For a
better understanding of the physical process, using Eq.~(\ref{dipi})
and its derivative for two dipolar objects, we can analytically
determine, through Eq.~(\ref{forcec}), the force on the spheres. Then,
on using the fact that there is a plane wave in the $Z$-direction (the
incident wave is $E_{0_{i}}e^{ik_0 z}$, where $i=x$ or $y$ depending
of the polarization of the incident field) the force on the second
sphere can be written as: \be\label{forcepolp} F_x(\ve{r}_2) & = & 1/2
\Re e\left(\alpha_2 E_i(\ve{r}_2,\omega) \alpha_1^*
E^*_i(\ve{r}_1,\omega) \frac{\partial}{\partial
x}T^*_{ii}(\ve{r}_2,\ve{r}_1,\omega) \right)\nonumber \\ & & \ee where
$i=x$ for $p$-polarization and $i=y$ for $s$-polarization of the
incident field. Notice that to obtain the force on sphere A, the
indices 1 and 2 must be permuted. But, even in this simple case, the
exact analytical solution of Eq.~(\ref{forcepolp}) is not easy to
interpret. Hence, we make in Eq.~(\ref{dipi}) the approximation that
the term $\ve{T}(\ve{r}^k_i,\ve{r}^l_j,\omega) \alpha^l_j$ is smaller
than 1 (we will discuss further this approximation).  Now, if we used
the hypothesis that the two spheres are identical
($\alpha_1=\alpha_2$), Eq.~(\ref{forcepolp}) becomes~: \be
\label{forcepolsimple} F_x(\ve{r}_2)=1/2|\alpha_1E_{0_i}|^2 
\Re e\left ( \frac{\partial}{\partial
x}T_{ii}(\ve{r}_2,\ve{r}_1,\omega)\right).\ee At short distance we can
make the non retarded approximation ($k_0=0$) and, as shown in the
appendix, we have that $F_x(\ve{r}_2)=-3|\alpha_1 E_{0_x}|^2/a^4$ in
$p$-polarization and $F_x(\ve{r}_2)=(3/2)|\alpha_1 E_{0_y}|^2/a^4$ in
$s$-polarization. The points (with the symbols $+$ and $\times$)
obtained with this approximation are shown in Fig.~2a and fit
correctly the curves obtained without any approximation, as seen in
this figure. Thus, they validate the approximation
$\ve{T}(\ve{r}^k_i,\ve{r}^l_j,\omega) \alpha^l_j\ll 1$ previously
done. Only when the spheres are very close to each other this
approximation slightly departs from the exact calculation due to the
increase of the free space susceptibility. In fact, this approximation
assumes the dipole associated to the spheres as only due to the
incident field, which is a good assumption when the polarizabilities
are small, like for glass spheres. It is now easy to physically
understand from Eq.~(\ref{forcepolsimple}) the reason of this either
attractive or repulsive force. As the spheres are small, the
scattering force is negligible~\cite{note} and thus only the gradient
force remains due to the interaction between the dipole associated to
sphere B, and to the variation of the field created by sphere A at the
position of sphere B. In $p$-polarization, the field due to sphere A
at the position of sphere B and the dipole of sphere B are in phase,
hence sphere B is pushed to the higher intensity region, namely
towards sphere A. In $s$-polarization, as the field due to sphere A at
the position of sphere B and the dipole of sphere B are in opposition
of phase, sphere B is pushed to the lower intensity region, namely,
far from sphere A. One can observe a similar effect in an atom
mirror~\cite{aspect}, or on a small silver particle in an evanescent
field.~\cite{prbchaumetnieto2}

On the other hand, in the far field we obtain, from the appendix, the
force upon sphere B as: $F_x(\ve{r}_2)=|\alpha_1 E_{0_x}|^2k_0^2
\cos(k_0 a)/a^2$ in $p$-polarization and $F_x(\ve{r}_2)=-|\alpha_1
E_{0_y}|^2k_0^3\sin(k_0 a)/(2a)$ in $s$-polarization, with
$k_0=2\pi\sqrt{\varepsilon_w}/\lambda$. The same explanation as before
can be used for the sign of the force: following the phase
relationship between the dipole and the field due to sphere A, the
force is either positive or negative, hence, the oscillations of the
force $F_x$ take place with period $\lambda/\sqrt{\varepsilon_w}$.
The phase difference $\lambda/(4\sqrt{\varepsilon_w})$ which appears
in the far field between the oscillations of $s$ and $p$ polarization,
comes from the difference between the derivative in the components
$xx$ and $yy$ of the free space susceptibility. We observe that the
force in $p$-polarization decreases faster than in $s$-polarization,
this is due to the absence of a propagating field along the $X$-axis
in the far field.  The magnitude of the force differs by a factor
$10^{4}$ between far field and near field. This is due to the strong
interaction between the spheres through the evanescent waves.

We can make an analogy in the near field with molecular physics. If we
look at the dipole moment of the two spheres, we compare our system of
forces with the interaction between two molecules. In
$p$-polarization, as the dipole moments are aligned and antisymmetric,
they produce an attractive force analogous to that between two
orbitals $p_z$, giving rise to a bonding state $\sigma_u$. In
$s$-polarization, the dipole moments are parallel and symmetric, so we
have antibonding states $\pi^*_g$, where $*$ means that the two
spheres cannot be bound.

\begin{figure}[H]  
\begin{center}
\includegraphics*[draft=false,width=80mm]{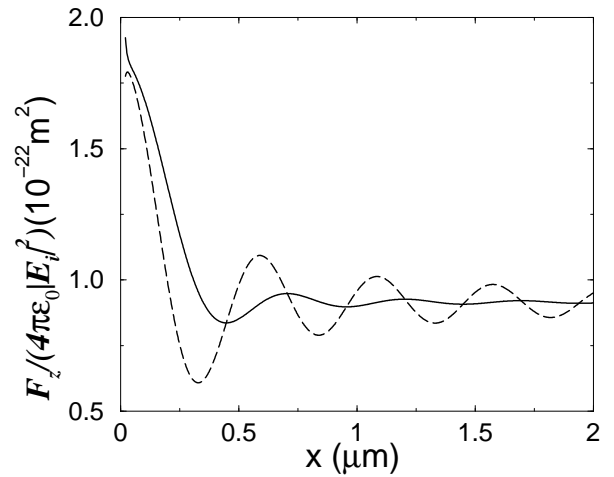}
\end{center}
\caption{Normalized force in the $Z$-direction on sphere B versus
distance $x$ between the centers of the spheres.  Both spheres are of
glass with $r=10$nm, $\theta=0^{\circ}$ and $\lambda=632.8$nm in
vacuum. The full line corresponds to $p$-polarization, whereas the
dashed line represents $s$-polarization.}
\end{figure}

We represent in Fig.~3 the force along the $Z$-direction. In this case
the scattering force is predominant.  The interaction between the
spheres is now directly responsible for the oscillation of the
force. Notice that when the spheres are far from each other, as the
interaction between the spheres becomes weak when the distance
increases, the force tends toward the scattering force upon one sphere
due to the incident field. As this force is not responsible of optical
binding, we are not going to discuss it further.

\begin{figure}[H]  
\begin{center}
\includegraphics*[draft=false,width=80mm]{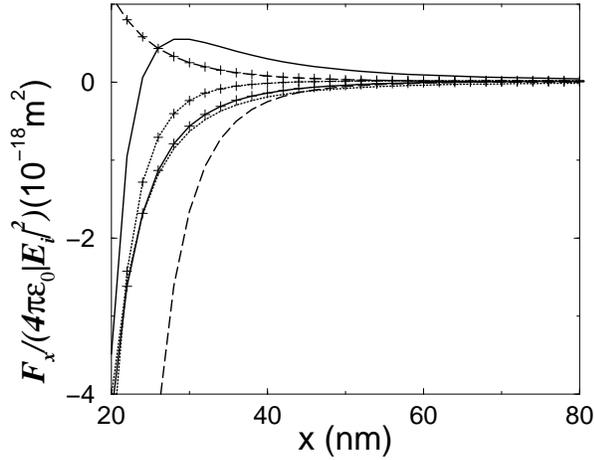}
\end{center}
\caption{Non-retarded approximation normalized force in the
$X$-direction on sphere B versus distance $x$ between the centers of
the spheres. The sphere A is of silver and the sphere B is of glass
with $r=$10nm. $\lambda=365$nm (full line), 388.5nm (dotted line), and
600nm (dashed line). Without symbol: $p$-polarization; with symbol +:
$s$-polarization.}
\end{figure}

More interesting is the case of two different small spheres, one (B)
being dielectric and the other (A) being metallic (silver). The first
fact easily observed from Eq.~(\ref{forcepolp}) pertains to the non
retarded case, as the derivative of the free space susceptibility is
real, the forces on spheres A and B are equal but of opposite sign to
each other (this is no longer the case in the far field). In Fig.~4 we
represent the force on sphere B at short distances from sphere A (in
comparison with the wavelength), for $\lambda=365$nm, 388.5nm, and
600nm. For $p$-polarization, we observe that the force at the
wavelength $\lambda=365$nm has a rather strange behavior as it is
positive, and only when the spheres are almost in contact, this force
changes and becomes similar to those at the other wavelengths. With
the approximation $\ve{T}(\ve{r}^k_i,\ve{r}^l_j,\omega) \alpha^l_j\ll
1$, the force in the non retarded approximation can be written \be
F_x(\ve{r}_2) & = & 1/2\frac{\partial
T_{ii}(\ve{r}_2,\ve{r}_1,\omega)} {\partial x}|E_{0_i}|^2 \alpha_2\Re
e(\alpha_1). \ee We observe that the sign of the force
depends of $\Re e(\alpha_1)$. When $\Re e(\alpha_1)>0$
($\lambda=600$nm), which is the common case, the dipole associated to
the silver sphere is in phase with the applied field, so everything
happens as for the dielectric sphere. Conversely, when $\Re
e(\alpha_1)<0$ ($\lambda=365$nm) the dipole is in opposition of phase
with the applied field, and hence the force becomes positive. But when
the spheres are almost in contact, the approximation
$\ve{T}(\ve{r}^k_i,\ve{r}^l_j,\omega) \alpha^l_j\ll 1$ is no longer
valid as shown when $\Re e(\alpha_1)=0$ ($\lambda=388.5$nm), case in
which the force is not null but negative. This is due to the fact that
the polarizability of the silver sphere is large and hence the
approximation is no longer valid for short distances. Physically, this
represents the contribution of the metallic sphere to the electric
field acting on the dielectric sphere, which is larger than that of
the incident field, hence the dipoles associated to the two spheres
are in phase and the force is attractive.  Notice that the change of
sign occurs both at the plasmon resonance ($\Re
e(\varepsilon_1)\approx -2\varepsilon_w$) and when $\Re
e(\varepsilon_1) \approx \varepsilon_w$. When $\varepsilon_1$ is
between these two values, the real part of the polarizability is
negative. For a more complete discussion on this, one can see
Ref.~[\ref{prbchaumetnieto2}]. A similar reasoning can be done for
$s$-polarization. If we now make the analogy previously done with the
molecular orbitals, then depending on the wavelength in
$s$-polarization, we shall obtain either antibonding states $\pi_g^*$,
or bonding states $\pi_u$.

In the far field, for $p$-polarization, on using Eq.~(\ref{forcepolp})
and the approximation $\ve{T}(\ve{r}^k_i,\ve{r}^l_j,\omega)
\alpha^l_j\ll 1$, we can write the force on the sphere B as \be
F_x(\ve{r}_2) & = & (\Re e(\alpha_1^*\alpha_2) \cos(k_0 a)\nonumber \\
& - & \Im m(\alpha_1^*\alpha2)\sin(k_0 a))k_0^2 |E_{0_x}|^2/a^2 \ee
and the force on sphere A as: \be F_x(\ve{r}_1) & = & (-\Re
e(\alpha_1^*\alpha_2) \cos(k_0 a)\nonumber\\ & - & \Im
m(\alpha_1^*\alpha_2)\sin(k_0 a))k_0^2 |E_{0_x}|^2/a^2 \ee As the
spheres are small, we can take only the gradient force as this is now
the predominant one, then $\alpha_2$ is real.~\cite{oplchaumetnieto}
Therefore, the forces on spheres A and B for the wavelength
$\lambda=600$nm, where $\Im m(\alpha_1)$ is weak, are opposite to each
other like for two identical spheres. But at $\lambda=388.5$nm, where
$\Re e(\alpha_1)=0$, the forces on the two spheres are completely
identical.

\begin{figure}[H]  
\begin{center}
\includegraphics*[draft=false,width=80mm]{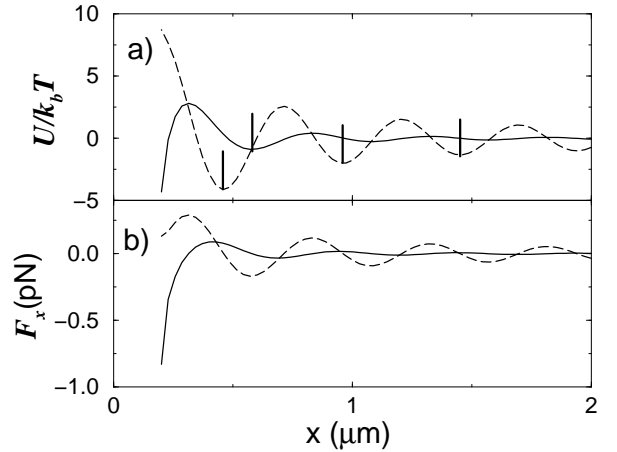}
\end{center}
\caption{Two glass spheres of radius $r$=100nm, $\theta=0^{\circ}$,
and $\lambda=632.8$nm in vacuum. The laser intensity of the incident
light is 0.2W/$\mu$m$^2$. Full line curves are for $p$-polarization.
Dashed line curves correspond to $s$-polarization. a) Potential of
sphere B normalized to $k_bT$ versus distance between the centers of
the spheres. The height of the bars correspond to a normalized
potential equal to 3.  b) Force in the $X$-direction versus the
distance between the spheres.}
\end{figure}

It should be remarked that if the laser intensity of the incident
light is assumed to be 0.2W/$\mu$m$^2$,~\cite{okamoto} the optical
forces for these small spheres are not strong enough to create an
optical binding, since then the Brownian motion stays the dominant
force. In this respect, the interest of the case of these small
spheres is mainly the interpretative value it yields of the underlying
physics. However, for larger radius in comparison to the wavelength,
the forces become larger and so is the trapping potential. In Figs.~5b
and~6b we plot the force along the $X$-axis for two dielectric spheres
(glass) with radius $r=100$nm, and 200nm, respectively. We observe
that with the intensity used previously (0.2W/$\mu$m$^2$) the
magnitude of the force is now enough to optically binding both
spheres. We compute the potential energy of the optical trap by
integration of the force (we take the potential energy null when the
second sphere is at infinity). As the two spheres are identical, the
potential energy is the same for both. The efficiency of the trapping
force requires it to be larger than the force due to the Brownian
motion, hence the depth of the potential wells of the trap should be
larger than $k_b T$, $T$ being the temperature of water, and $k_b$ the
Boltzmann constant. Considering $T=$290K, then $k_bT=4\times
10^{-21}$J.  We plot in Figs.~5a and~6a the potential normalized to
the value $k_bT$. We adopt the criterion that the trap is efficient
when the potential well is larger that 3$k_b T$. Hence the bars
plotted at the bottom of the wells in Figs.~5a and~6a correspond to
the value 3.  We see from Fig.~5 that for $p$-polarization the trap is
not feasible except when the spheres are in contact. For
$s$-polarization, we have three equilibrium positions spaced out by
one wavelength. This behavior is explained by the previous results on
small spheres, and in agreement with experiments.~\cite{burns} When
the size of the sphere is close to one wavelength, we see from Fig.~6
that the depth of the potential well is larger than in the previous
case. In $p$-polarization there is no possibility to stick the spheres
together, but now we have one potential minimum of stable
position. Thus, we observe that it is easier to trap particles when
their radius are large, in agreement with the experiments of Burns et
al.~\cite{burns}.

\begin{figure}[H]  
\begin{center}
\includegraphics*[draft=false,width=80mm]{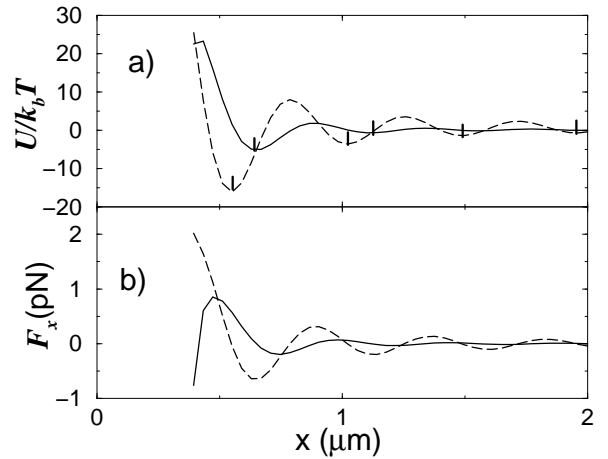}
\end{center}
\caption{Same as Fig.~5 but with a radius $r=$200nm.}
\end{figure}

Notice that the gravity force is $6.16\times 10^{-5}$ pN and
$4.93\times 10^{-4}$ pN for the spheres of radii r=100nm and 200nm,
respectively. Another force that exists between the spheres is the
Casimir-Polder force, $F_c$.  To our knowledge, $F_c$ has often been
studied either between two plates or between a sphere and a
plate,~\cite{klimchitskaya} but it has never been established between
two spheres. In the non retarded approximation and the dipole
approximation for the spheres, $F_c$ is reduced to the dispersion
force (London's force) which is inversely proportional to the seventh
power of the distance between the dipoles (see for example
Ref.~[\ref{milonni}]). Hence, only when the spheres are in contact, or
at distances smaller than the wavelength from each other, this force
might be of the same magnitude of the optical forces. However, the
fast decay of this force at distances larger than the wavelength
prevents it to perturb the optical trap.

In table~1 we give some examples on the limiting radius to get optical
trapping for two identical spheres embedded in water (notice that for
$p$ polarization we do not mean optical trapping when the spheres are
sticked in contact), i.e., the minimum radius to obtain one potential
minimum of stable position for the two spheres using the same
criterion as before (namely, $U>3k_bT$).

\end{multicols}

\begin{table}
\begin{center}
\begin{tabular}{|c|c|c|c|}
spheres & glass ($\lambda=632.8$nm) & silver ($\lambda=394$nm)
& silver ($\lambda=314$nm)\\ \hline limiting radius ($p$-pol.) & 123 nm
& 33 nm & 180 nm \\ \hline limiting radius ($s$-pol.) & 85 nm & 21 nm &
50 nm \\ 
\end{tabular} 
\end{center}  
\caption{Minimum radius to get one minimum position of the potential
for two identical spheres. The following cases are addressed: glass
sphere, silver sphere (both off and in plasmon resonance) The
criterion of stability used is: the potential well depth must be
larger that 3$k_b T$.}
\end{table}

\begin{multicols}{2}

We should remark that this is the limiting radius only to obtain the
first stable position, if we want to get more stable positions, as in
Ref.[\ref{burns}], the radius must be larger.  As mentioned before,
the table shows that the optical trapping is easier for $s$
polarization. For the silver sphere, the value $\lambda=394$nm
corresponds to the plasmon resonance, and that of $\lambda=314$nm is
for a wavelength out of resonance. At the plasmon resonance the
polarizability is largest, so it is easier to perform optical binding
at this wavelength.

\begin{figure}[H]  
\begin{center}
\includegraphics*[draft=false,width=80mm]{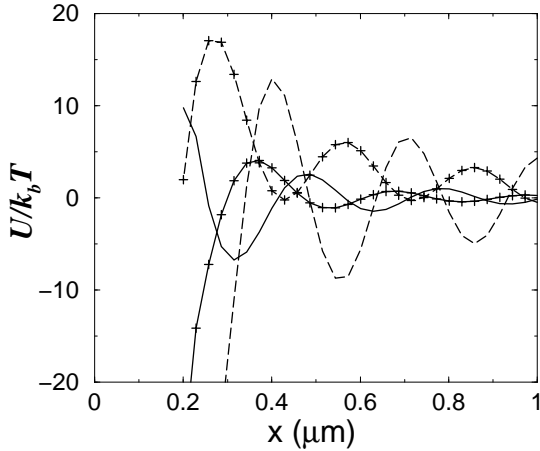}
\end{center}
\caption{The sphere A is of glass and the sphere B of silver with
$r=$100nm, $\lambda=388.5$nm. The laser intensity of the incident
light is 0.2W/$\mu$m$^2$. Plot of the potential normalized to $k_bT$
for the two spheres versus the distance between them. The full line is
for $p$-polarization and dashed line corresponds to
$s$-polarization. The potential of sphere A is without symbol, and the
potential of sphere B is with symbol $+$.}
\end{figure}

As a last instance, we now consider two spheres in free space, with
radii $r=$100nm, one being of glass and the other of silver,
illuminated by a plane wave at $\lambda=$388.5nm. We plot in Fig.~7
the potential energy for the two polarizations and the two spheres,
since now this magnitude depends on the sphere material. We then
observe that it is not possible to obtain a stable equilibrium since
the potentials of the two spheres are now different. This result is
explainable from the previous calculation on small spheres (of silver
and glass), since then the forces were always opposed to each other at
this wavelength. In fact, as the spheres are now large, the forces at
this wavelength are not exactly in opposition, due to the larger
scattering and absorbing force.  Hence it is possible to obtain points
where the potential of the two spheres is minimum. This happens when
the forces on each sphere are the same and positive. In that case, the
two spheres move in the direction of the positive $X$-axis while
keeping constant the distance between them.

\subsection{Particles in water on a dielectric flat surface}\label{resultsB}

In this section we consider a flat dielectric surface upon which the
spheres are suspended in water, as shown by Fig.~1. We compute the
force along the $X$-axis on sphere B when both spheres are dielectric
(glass), with $\theta=0^{\circ}$ (Fig.~8).  We now observe that both
for large and small spheres the force has a behavior similar to that
acting on dielectric spheres isolated in water. When the spheres are
in contact, the force on sphere B is the same as in absence of
interface, whereas when the spheres are far from each other, this
force is slightly smaller than without interface.  This means that the
optical binding is more difficult to perform when the spheres are on a
surface than when they are far from interfaces. Also, there is a
change in the period of oscillation due to interaction between the
spheres via the light reflected by the surface.  However, in the case
when one of the spheres is metallic (silver), we observe the same
behavior of the forces when the surface is present as without it.
Then, as previously observed for dielectric spheres, there only
appears a shift in the oscillation and magnitude of the forces.

\begin{figure}[H]  
\begin{center}
\includegraphics*[draft=false,width=80mm]{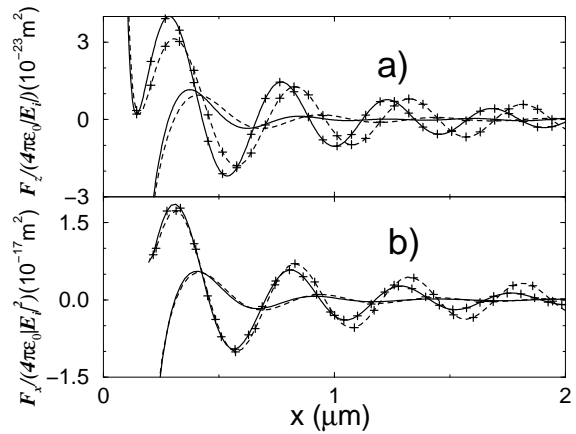}
\end{center}
\caption{Force in the $X$-direction upon sphere B when the spheres are
placed on a flat dielectric surface.  $\theta=0^{\circ}$,
$\lambda=632.8$nm. The full line represents the force in the presence
of the surface, and the dashed line corresponds to the force computed
without the interface. The curves with symbols + denote
$s$-polarization, and those without symbol correspond to
$p$-polarization. a) spheres of glass with $r=$10nm. b) spheres of
glass with $r=$100nm.}
\end{figure}

In Fig.~9 we investigate the potential and optical force on two glass
spheres in front of a dielectric surface, illuminated by total
internal reflection ($\theta=50^{\circ}$).  Figure~9b shows that the
force component along the $X$-axis always pushes the spheres in the
direction of the wave vector component parallel to the surface. Hence,
it is not possible to obtain a stable equilibrium with the two spheres
remaining fixed. But if we compute the potential of the two spheres
together (Fig.~9a), we observe some minima indicating that the system
can acquire internal equilibrium, namely, the relative positions of
the spheres can be kept fixed. Hence, when both spheres move impelled
by the evanescent wave propagating along the surface, their velocity
remains parallel to this surface, while the distance between them
keeps some particular values given by the position of the potential
minima (cf. Fig.~9a). Notice that the force on the second sphere (in
both polarizations) has no oscillation, a very similar behavior was
observed by Okamoto et {\it .al}.~\cite{okamoto}. The computational
prediction of similar collective movements in systems of more than two
spheres will involve long computing times of their relative positions
by potential energy minimization.

\section{Conclusion}

We have studied the optical binding between two spheres embedded in
water, either in presence or absence of a flat dielectric
interface. We have presented results for different sizes and
illumination conditions.  Some of them agree with previous
experiments~\cite{burns} for two identical spheres, however, when they
are composed of different material, the force between them may have
quite different behavior, depending on the light wavelength
employed. In future work, it would be interesting to investigate the
effect of light on several spheres in water in order to build up
particle arrays. However, in this respect, the vertical force that
pushes the spheres away from the substrate, constitutes a hindrance to
this aim. This work shows, however, that this problem can be avoided
by illuminating the system under total internal reflection at the
substrate interface. Then the spheres will be sticked to the surface
by the gradient force due to the transmitted evanescent wave. The
horizontal force of this surface wave on the spheres that pushes them
along the interface, can be compensated by means of a second
counterpropagating evanescent wave, created by an additional
beam. Notice, in addition, that if both surface waves are mutually
coherent, the resulting standing wave pattern can introduce further
structure in the resulting potential wells acting on the spheres.

\begin{figure}[H]  
\begin{center}
\includegraphics*[draft=false,width=80mm]{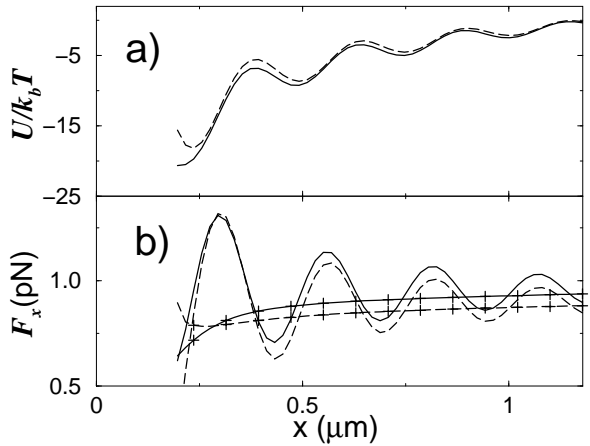}
\end{center}
\caption{Fig.~9 Two glass spheres of radius $r$=100nm in vacuum, in
 front of a flat dielectric surface. $\theta=50^{\circ}$,
 $\lambda=632.8$nm. The laser intensity of the incident light is
 0.2W/$\mu$m$^2$. Full line curves are for $p$-polarization.  Dashed
 line curves correspond to $s$-polarization. a) Potential of
 interaction between the two spheres, normalized to $k_bT$, versus
 distance between the centers of the spheres. b) Force in the
 $X$-direction against distance between the spheres. The force curve
 on sphere A is without symbol, and the force on sphere B is with
 symbol $+$.}
\end{figure}

\section{Acknowledgements}

Work supported by the European Union and the Direccion General de 
Investigacion Cintifica y Tecnica, grant PB98 0464.

\appendix

\section{Derivative of the free space susceptibility}

The derivative of the free space susceptibility used in this paper is:
\be \frac{\partial}{\partial x}T_{xx}(x,x_0)=-\frac{6\ve{a}}{a^5},\\
\frac{\partial}{\partial x}T_{yy}(x,x_0)=\frac{\partial}{\partial
x}T_{zz}(x,x_0)=\frac{3\ve{a}}{a^5},\ee in the non retarded case, and
\be \frac{\partial}{\partial
x}T_{xx}(x,x_0,\omega)=\frac{2\ve{a}k_0^2}{a^3}e^{ik_0 a},\\
\frac{\partial}{\partial
x}T_{yy}(x,x_0,\omega)=\frac{\partial}{\partial
x}T_{zz}(x,x_0,\omega)=\frac{i\ve{a}k_0^3}{a^2}e^{ik_0 a},\ee in the
far field, where $\ve{a}=(x-x_0)$ and $a=|\ve{a}|$. $x$ is the
abscissa of the observation point, and $x_0$ that of dipole position.


\end{multicols}

\end{document}

%% file: fig1.pstex_t
\begin{picture}(0,0)%
\epsfig{file=fig1.pstex}%
\end{picture}%
\setlength{\unitlength}{3947sp}%
\begingroup\makeatletter\ifx\SetFigFont\undefined%
\gdef\SetFigFont#1#2#3#4#5{%
  \reset@font\fontsize{#1}{#2pt}%
  \fontfamily{#3}\fontseries{#4}\fontshape{#5}%
  \selectfont}%
\fi\endgroup%
\begin{picture}(5274,4242)(2989,-4723)
\put(5251,-2011){\makebox(0,0)[lb]{\smash{\SetFigFont{20}{24.0}{\familydefault}{\mddefault}{\updefault}$a$}}}
\put(3526,-4486){\makebox(0,0)[lb]{\smash{\SetFigFont{20}{24.0}{\familydefault}{\mddefault}{\updefault}$\theta$}}}
\put(4426,-3211){\makebox(0,0)[lb]{\smash{\SetFigFont{20}{24.0}{\familydefault}{\mddefault}{\updefault}$O$}}}
\put(3796,-766){\makebox(0,0)[lb]{\smash{\SetFigFont{20}{24.0}{\familydefault}{\mddefault}{\updefault}$Z$}}}
\put(7711,-3286){\makebox(0,0)[lb]{\smash{\SetFigFont{20}{24.0}{\familydefault}{\mddefault}{\updefault}$X$}}}
\put(6526,-1861){\makebox(0,0)[lb]{\smash{\SetFigFont{20}{24.0}{\familydefault}{\mddefault}{\updefault}B}}}
\put(4051,-1861){\makebox(0,0)[lb]{\smash{\SetFigFont{20}{24.0}{\familydefault}{\mddefault}{\updefault}A}}}
\put(6376,-2611){\makebox(0,0)[lb]{\smash{\SetFigFont{20}{24.0}{\familydefault}{\mddefault}{\updefault}$\varepsilon_2$}}}
\put(6901,-2086){\makebox(0,0)[lb]{\smash{\SetFigFont{20}{24.0}{\familydefault}{\mddefault}{\updefault}$r$}}}
\put(3676,-2311){\makebox(0,0)[lb]{\smash{\SetFigFont{20}{24.0}{\familydefault}{\mddefault}{\updefault}$\varepsilon_1$}}}
\put(4876,-1111){\makebox(0,0)[lb]{\smash{\SetFigFont{20}{24.0}{\familydefault}{\mddefault}{\updefault}$\varepsilon_w=1.69$}}}
\put(4951,-3886){\makebox(0,0)[lb]{\smash{\SetFigFont{20}{24.0}{\familydefault}{\mddefault}{\updefault}$\varepsilon_s=2.25$}}}
\put(3001,-3886){\makebox(0,0)[lb]{\smash{\SetFigFont{20}{24.0}{\familydefault}{\mddefault}{\updefault}$\ve{k}_0$}}}
\end{picture}

%% file: optical_bind.bbl
\begin{references}
\bibitem{note} For a small sphere, the gradient and scattering forces are 
proportional to $a^3$ and $a^6$, respectively.
\end{references}
